\documentclass[11pt]{article}
\usepackage{moriond,epsfig}
\usepackage{xspace}
\usepackage{subfig}

\def\Journal#1#2#3#4{{#1} {\bf #2}, #3 (#4)}
\def\D0{D\O\xspace}
\def\GeVcsq{\ensuremath{{\rm GeV}/c^2}\xspace}
\def\TeVcsq{\ensuremath{{\rm TeV}/c^2}\xspace}
\def\invfb{\ensuremath{{\rm fb}^{-1}}\xspace}
\def\MET{\ensuremath{{\not\!\! E}_{T}}\xspace}

\def\HT{\ensuremath{H_{T}}\xspace}
\def\LTP{\ensuremath{\rm{LTP}}\xspace}

\def\PLB{{\em Phys. Lett.}  B}
\def\PRL{\em Phys. Rev. Lett.}
\def\PRD{{\em Phys. Rev.} D}

\def\be{\begin{equation}}
\def\ee{\end{equation}}
\def\bea{\begin{eqnarray}}
\def\eea{\end{eqnarray}}

\begin{document}
\vspace*{4cm}
\title{NON-SUSY SEARCHES AT THE TEVATRON}

\author{ANTONIO BOVEIA\footnote{on behalf of the CDF and \D0 Collaborations.}}

\address{Enrico Fermi Institute, University of Chicago$\:$\footnote{formerly of the University of California, Santa Barbara.},\\
5640 S Ellis Ave, Chicago, IL 60637, USA}

\maketitle\abstracts{
  I describe recent signature-based searches for anomalous physics processes with up to 2.9 \invfb of data from the CDF or \D0 detectors at the Fermilab Tevatron. While each search conveys its sensitivity by interpreting a null result in terms of one or more specific exotic models, the searches are designed to be broadly sensitive to many models.
}

\section{Introduction}
Without unambiguously new phenomena to explore or a truly compelling Standard Model extension to test, the highest-energy collider experiments often look for anomalous physics via model-independent ``signatures'' which are general enough to probe a broad variety of models but specific enough to allow sufficient optimization against backgrounds. Along with targeted analyses, such as the searches for supersymmetry discussed elsewhere in these proceedings, CDF and \D0 perform many signature-based searches. I discuss some recent results here; searches involving top quarks or signature-independence are described in separate contributions.

\section{Searches for dilepton resonances}
\begin{figure} \centering
\includegraphics[height=2in]{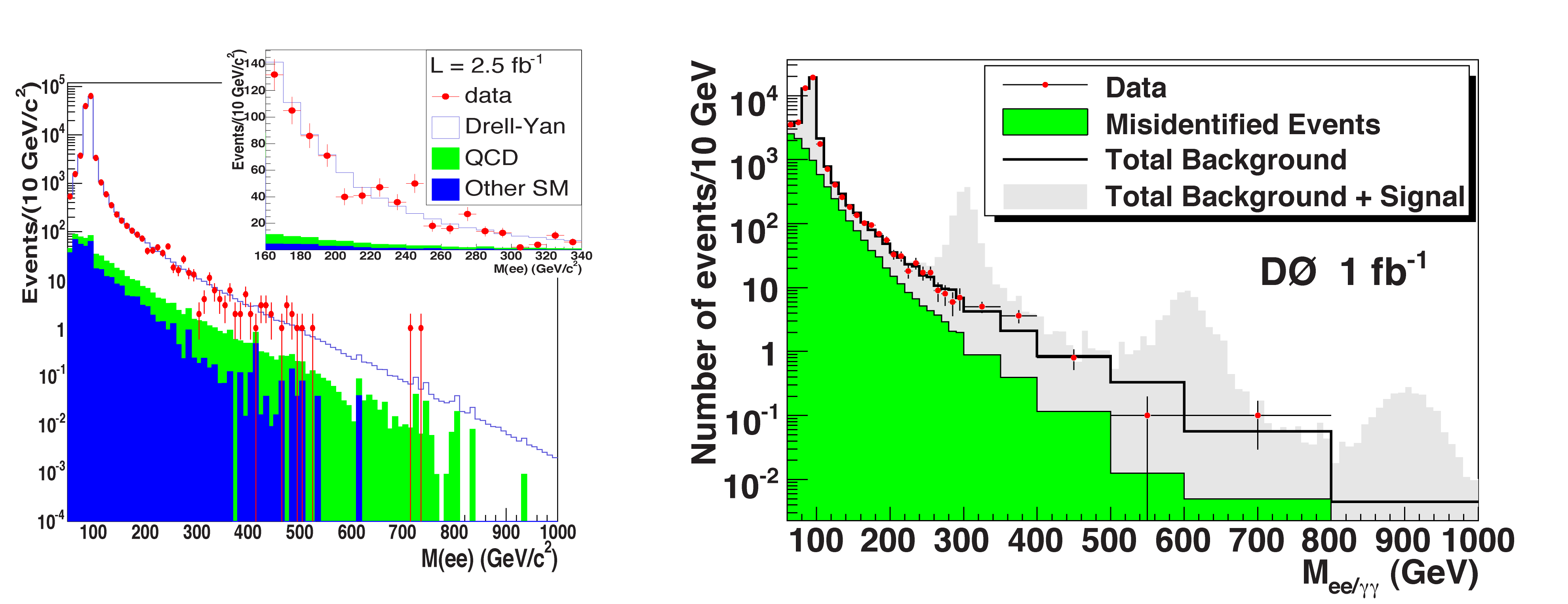}\caption{CDF dielectron invariant mass (left) and \D0 di-EM (electron/photon) invariant mass (right).\label{fig:mee-spectra}}
\end{figure}

$Z'$ searches for a resonant excess in the invariant mass distribution of two leptons are an ideal signature-based search---their distinguishing feature is simple; their final states and backgrounds are well understood both experimentally and theoretically; and their reach is far, addressing many popular proposals such as Randall-Sundrum (RS), technicolor, Little Higgs, E6, and R-parity violating supersymmetry models\cite{carena}. Both CDF and \D0 have looked for $Z'$ resonances in the dielectron mass spectrum above the $Z$ peak. When combined with a detector simulation and various small backgrounds, the Pythia Drell-Yan calculation correctly predicts the dielectron spectrum observed in 2.5 \invfb of CDF data\cite{cdfee}. Though the data contain an upward fluctuation at about 240 GeV, a fluctuation at least as large at a mass heavier than 150 GeV should occur in about 0.6\% of experiments (2.5$\sigma$). \D0 performs a similar search in the combined dielectron and diphoton channel with 1 \invfb and finds no comparable excess at any mass\cite{d0diem}. Both spectra are shown in Figure \ref{fig:mee-spectra}. CDF also has a search for $Z' \rightarrow \mu\mu$ with 2.3 \invfb of data. Since the tracker resolution in curvature is roughly constant with momentum, the search is done in inverse dimuon mass $1/M_{\mu\mu}$. The dimuon data again agree with the Drell-Yan--dominated Standard Model prediction\cite{cdfmm}.

One can communicate the sensitivity of these searches by setting limits on a theoretical model containing a massive dilepton resonance. At 95\% confidence, CDF excludes spin 1 SM-like $Z'$ bosons lighter than 965 \GeVcsq with the dielectron channel and lighter than 1.03 \TeVcsq with the dimuon channel. For RS1 scenarios with $k/M_p = 0.1$, \D0's result excludes virtual gravitons lighter than 900 \GeVcsq and the CDF dimuon result excludes gravitons lighter than 921 \GeVcsq. \D0 also provides constraints on ADD-like extra-dimensional models with a near-continuum of narrowly spaced Kaluza-Klein (KK) gravitons. For $n=2$ extra dimensions, $\bar{M_p} > 2.09\:\TeVcsq$; for $n=7$, $\bar{M_p} > 1.29\:\TeVcsq$.

\section{Searches for photon(s) or jet(s) and missing momentum}
Both experiments search in the single or double photon or jet and missing momentum channels. As with $Z'$ searches, these are experimentally and theoretically attractive signatures---in this case, of the production of a dark matter candidate in supersymmetry, continuum KK graviton emission, and other models. With 1.0 \invfb of data, the \D0 single photon + \MET analysis obtains a photon transverse momentum spectrum that agrees with the background prediction, a concoction of Z+$\gamma$ (irreducible when the Z decays to neutrinos) and instrumental backgrounds\cite{d0gmet}. A similar search for missing momentum and either a single photon or a single jet in 2 \invfb of CDF data again finds no evidence of anomalous production\cite{cdfgjmet}. These results constrain the higher-dimensional Planck mass, surpassing limits from LEP for $n \geq 4$ extra dimensions.

With 2 \invfb, CDF analyzes distributions of \MET and the scalar sum of photon transverse momenta (\HT) in the diphoton + \MET signature, a feature of GMSB supersymmetry and models with massive stable graviton production. At low \MET, mismeasured jets comprise the main background. To predict this background, the analysis uses a detailed parameterization of spurious \MET obtained from the recoil jets in $Z\rightarrow e^+ e^-$ events. At high \MET, the main backgrounds are electroweak processes involving intrinsic \MET, obtained from simulation. The result consists of counting experiments and \MET and \HT distributions for various minimum \MET significance requirements. No excess production is observed\cite{cdfggmet}.

\begin{figure} \centering
  \includegraphics[height=2.25in]{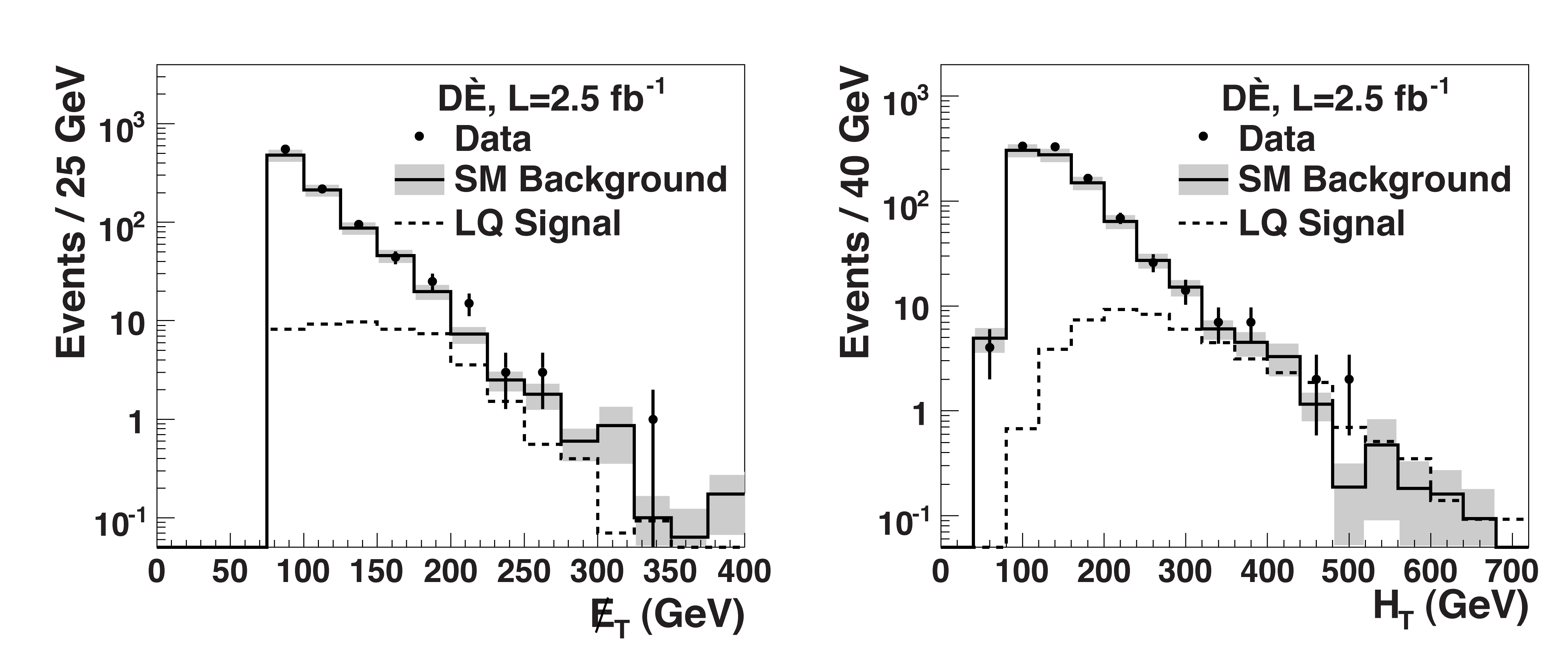}
  \caption{Key distributions for the \D0 dijet + \MET analysis, with the expected contribution from a 200 GeV scalar leptoquark.}
  \label{fig:dijetmet}
\end{figure}

\D0 analyzes 2.5 \invfb for the dijet + \MET signature using simultaneous requirements on the minimum \MET and the minimum scalar sum \HT of jet transverse momentum\cite{d0jjmet}. As shown in Figure \ref{fig:dijetmet}, both distributions agree with expectation in the bulk, SM-dominated region (predominantly $Z\rightarrow\nu\nu$ + jets production) and in the high tails where one might have seen anomalies. \D0 expresses this null result as limits on scalar leptoquarks ($M_{LQ} > 205\:\GeVcsq$ when $LQ$ decays exclusively to $j\nu$) and limits on the lightest T-odd particle in Little Higgs models (if $\tilde{Q}\rightarrow j\:\LTP$ exclusively and $M_{\LTP} < 200\:\GeVcsq$, then $M_{\tilde{Q}} >~ 400\:\GeVcsq$ at 95\% confidence).

\section{Searches for diboson resonances}
Diboson resonance searches are an experimentally attractive analog of the $Z'$ searches---they involve the same clean final states but also have additional particles or mass constraints which lead to very small backgrounds. \D0 has a search for a massive particle decaying to a $Z$ and a photon, where the $Z$ subsequently decays to electrons or muons\cite{d0zg}. With 1 \invfb of data, the dilepton + photon invariant mass spectrum agrees with the SM expectation, derived from a Baur calculation of $Z\gamma$ and a fake-rate--based estimate of the contribution from $Z$+jets. Without a clear signal, \D0 sets limits on particle production cross sections assuming the acceptance of either a scalar or a vector particle. For example, the search sets an upper limit of about 200 fb for 600 \GeVcsq particles of either type.

CDF also has new searches in heavy gauge boson modes, which can probe high mass higgs and bulk RS scenarios. One search involves a massive particle decaying to $e\nu jj$ through a diboson pair\cite{cdfww}. The $WW$ and $WZ$ invariant mass spectra agree with the SM prediction consisting of $W$ + jets at low mass and a concoction of backgrounds, such as QCD faking the trigger $W$, at high mass. The search interprets the null result as limits on W' ($284 < M_{W'} 515\:\GeVcsq$), Z' ($247 < M_{Z'} < 545\:\GeVcsq$), and RS graviton models ($M_{G} > 607\:\GeVcsq$).

\begin{figure} \centering
  \subfloat{\includegraphics[height=1.5in]{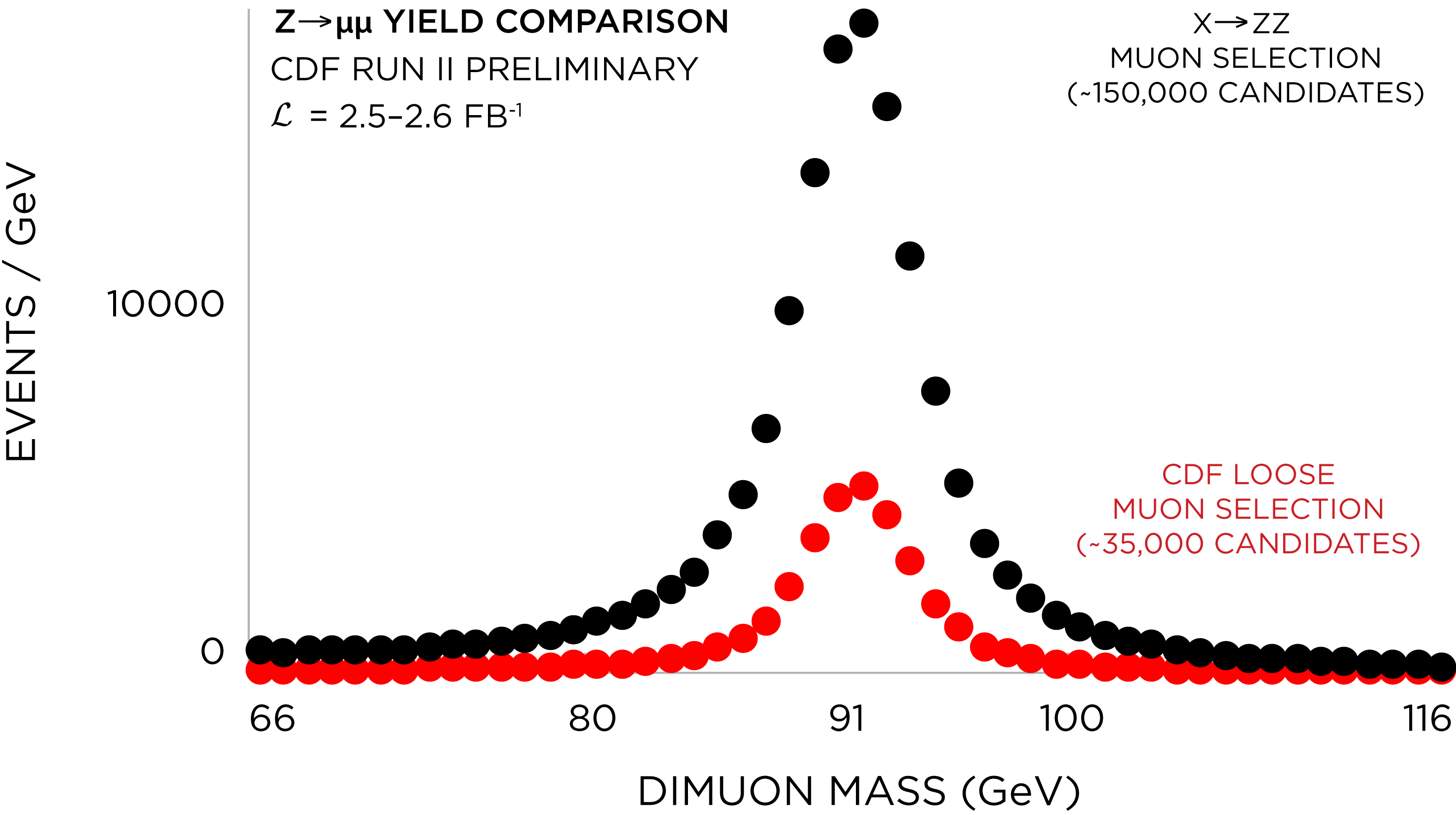}}\\
  \subfloat{\includegraphics[height=1.5in]{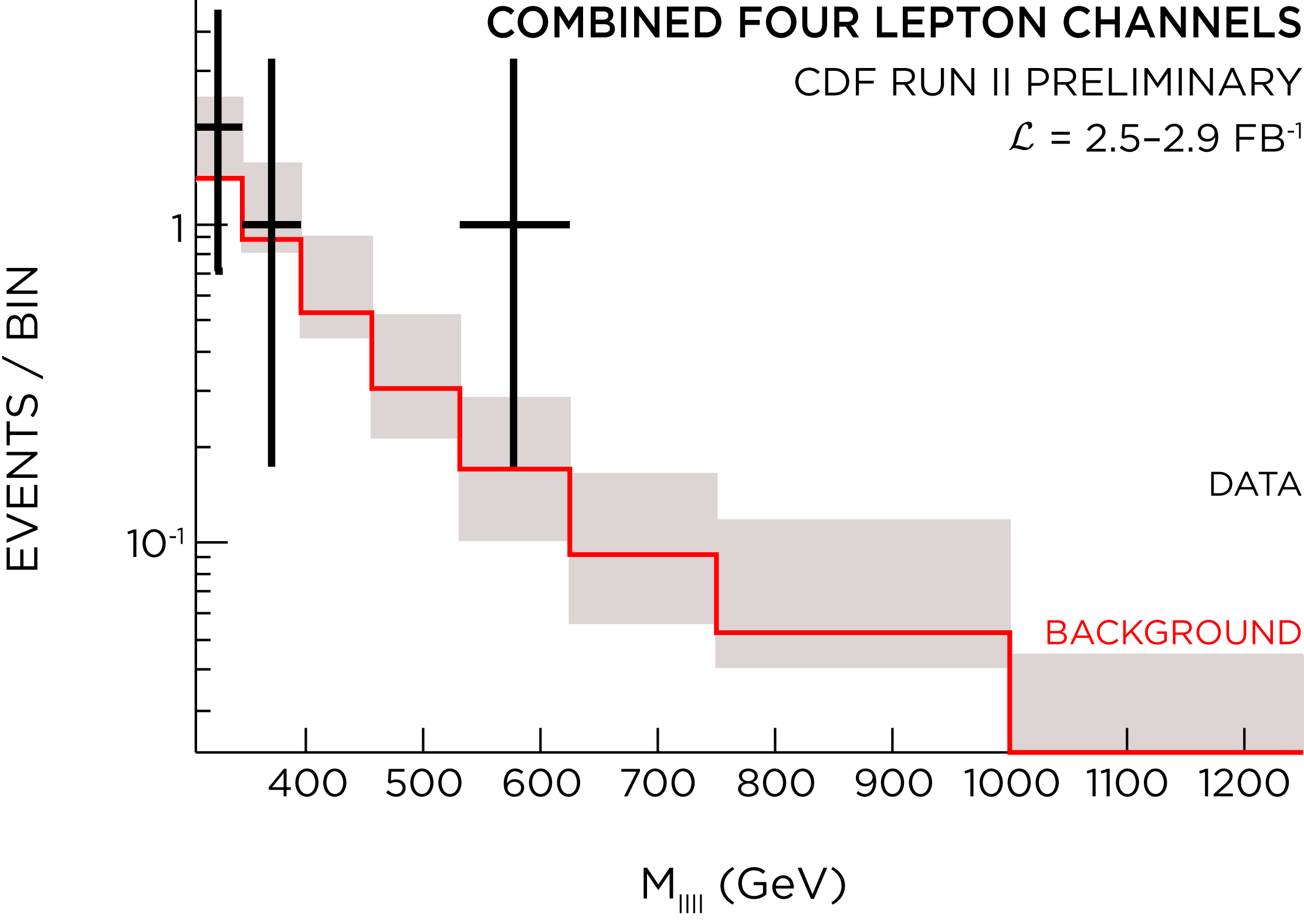}}\quad\quad\quad
  \subfloat{\includegraphics[height=1.5in]{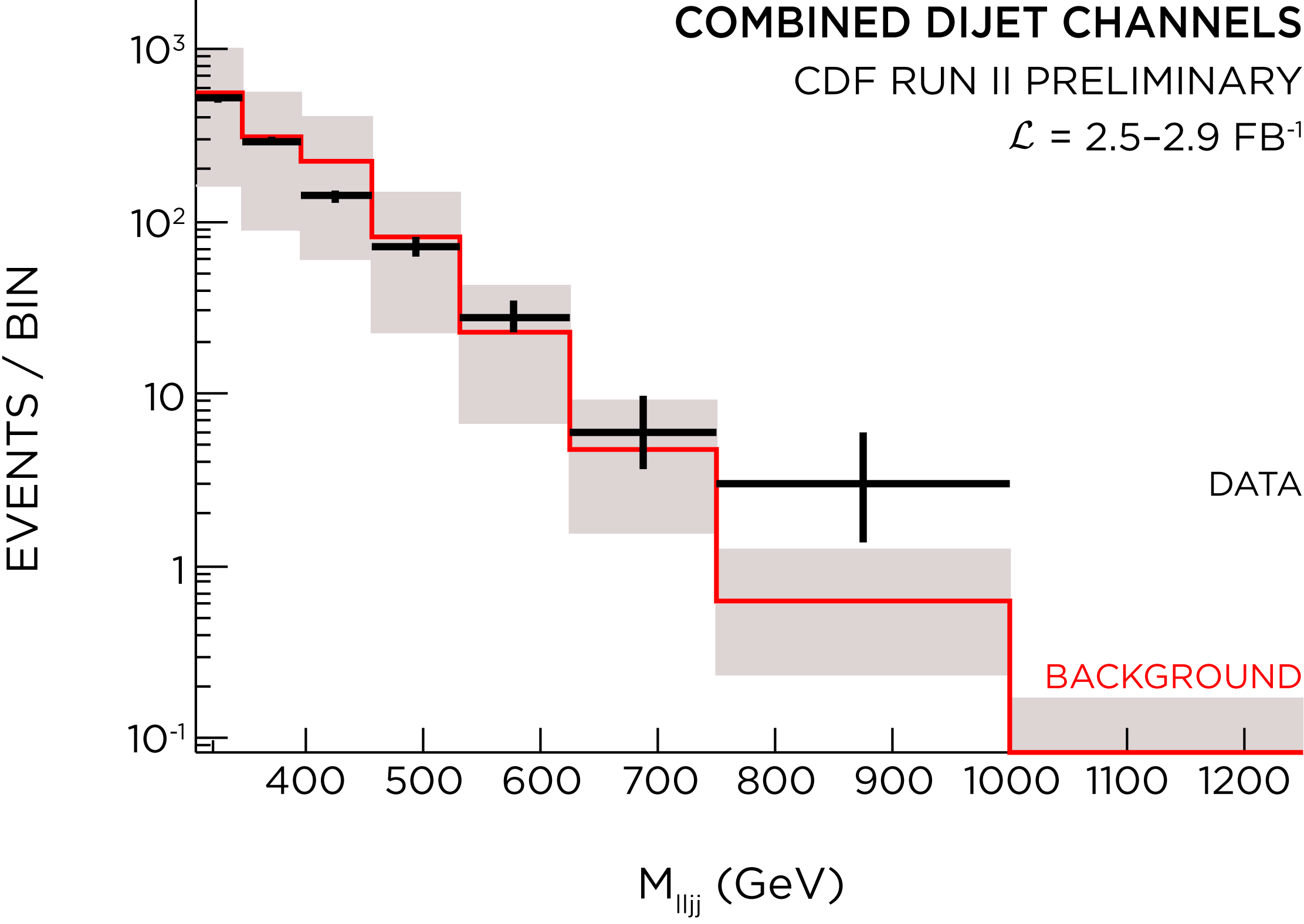}}
  \caption{$X \rightarrow ZZ$ analysis: (top) comparison of dimuon yields using standard and improved muon selections and (bottom) four body mass spectra and background predictions for the four lepton and two lepton, two jet channels ($l=e,\mu$).}\label{fig:xzz}
\end{figure}

CDF also has a result in the $X \rightarrow ZZ$ signature using dilepton + dijet decays and four lepton decays\cite{cdfzz}. This result is the first to use CDF's newly improved forward track reconstruction, and it also uses more efficient electron and muon selection than typically employed. The cumulative effect of these improvements is demonstrated with data by the dimuon mass spectrum, where relative to muon criteria used in other CDF analyses the improvements more than quadruple the yield in the $Z$ peak (Figure \ref{fig:xzz}). The $X \rightarrow ZZ$ results are the spectra of the four lepton invariant mass and the two lepton two jet invariant mass, also shown. The dominant backgrounds are not resonant in both $Z$ masses and are estimated entirely from data using sideband fits. The data agree well with the background prediction; there is no evidence of any bump in the mass spectra. The analysis conveys its sensitivity with a limit on RS graviton-like particles decaying to $ZZ$, ruling out masses below 491 \GeVcsq.

\clearpage

\section{Conclusions}

None of the signature-based searches discussed here uncovers a substantial population of anomalous events, but in many cases these null results do not yet preclude discovery. Both CDF and \D0 have recorded 1.8--5 times as much data. As the datasets continue to grow, Tevatron searches may yet find an interesting excess.


\section*{References}

\begin{thebibliography}{99}
\bibitem{carena}M. Carena, et al., \Journal{\PRD}{70}{93009}{2004}.
\bibitem{cdfee} CDF Conference Note 9160 (2008).
\bibitem{d0diem} \D0 Collaboration, \Journal{\PRL}{100}{91802}{2008}.
\bibitem{cdfmm} CDF Collaboration, \Journal{\PRL}{102}{91805}{2009}.
\bibitem{d0gmet} \D0 Collaboration, \Journal{\PRL}{101}{11601}{2008}.
\bibitem{cdfgjmet} CDF Collaboration, \Journal{\PRL}{101}{161602}{2008}.
\bibitem{cdfggmet} CDF Conference Note 9625 (2008).
\bibitem{d0jjmet} \D0 Collaboration, \Journal{PLB}{668}{357}{2008}.
\bibitem{d0zg} \D0 Collaboration, \Journal{\PLB}{671}{349}{2009}.
\bibitem{cdfww} CDF Conference Note 9730 (2009). 
\bibitem{cdfzz} CDF Conference Note 9640 (2008), FERMILAB-THESIS-2008-53.
\end{thebibliography}
\end{document}